\documentclass[preprint]{revtex4-2}

\usepackage{graphicx}
\usepackage{dcolumn}
\usepackage{bm}
\usepackage{amsmath}
\usepackage{tabularx}
\usepackage{caption}
\usepackage{subcaption}
\usepackage{natbib}

\begin{document}

\title{Impact of Topology and Sequence Length on the Pressure-Dependent Collapsibility of Proteins}
\date{\today}

\author{Jiayi Wang}
\affiliation{Department of Physics, University of Washington}

\author{D. Thirumalai}
\affiliation{Departments of Chemistry and Physics,
The University of Texas at Austin, Austin 78712}

\author{Margaret S. Cheung}
\email{margaret.cheung@pnnl.gov}
\affiliation{Department of Physics, University of Washington}
\affiliation{Pacific Northwest National Laboratory}

\author{Andrei G. Gasic}
\email{andrei@geminus.ai}
\affiliation{Geminus.ai}

\begin{abstract}
 Several physical and evolutionary factors, including temperature, pressure, solvent quality, and natural selection, play an important role in guiding amino acid sequences toward their native structures. Although the folding mechanisms of single-domain proteins have been extensively studied, the physical determinants of protein collapse and their relationship to native-state folding is much less understood. In particular, the notion of "collapsibility", the intrinsic propensity of a protein sequence to adopt compact conformations under folding conditions, is distinct from but closely related to "foldability", the ability to reach the native structure. Previous theoretical studies suggested that collapsibility is encoded in native-state topology, highlighting a potential role of natural selection in favoring sequences with a greater propensity to collapse. However, the full story of natural selection cannot be told without the impact of water and hydrostatic pressure. Here, we investigate the role of solvent-mediated interactions in protein collapse by incorporating a pressure-dependent desolvation potential and examine how hydrostatic pressure influences protein collapsibility. We find that, although increasing pressure destabilizes the fully folded native state, it can still favor the formation of compact, water-mediated conformations. In addition, the extent and direction of the pressure-induced response depend strongly on native-state topology, including both secondary and tertiary structures. Proteins or domains containing a higher proportion of non-local contacts, for example, $\beta$-sheet-rich proteins, generally exhibit a more pronounced tendency towards water-mediated collapse with increased pressure. These findings highlight the critical role of native-state topology in governing the pressure sensitivity of protein collapse and provide insights into the interplay between protein topology, solvent-mediated interactions, and conformational transitions in response to changes in the environment.

\end{abstract}
\maketitle

\section{\label{sec:level1}Introduction}
Although considerable progress has been made in predicting protein three-dimensional structures, understanding the physical mechanisms and pathways of protein folding remains a fundamental challenge \cite{energy_landscape, Onuchic04COSB, flory_original, Thirumalai_1, Thirumalai_2, Thirumalai_3, Thirum95JPI, baker_topology, gianni2021fuzziness, wang2024role}. Classical polymer theories, such as Flory theory, provide a framework for describing polymer conformations and thermodynamic properties by representing chains as interacting monomers \cite{flory_original}. However, proteins differ from homopolymers in their sequence-specific interactions and evolutionary constraints. Natural selection favors only a small subset of possible protein sequences and native structures \cite{isakova2026descent,hilser2025statistical}. Energy landscape theory explains this selection through minimally frustrated, funneled landscapes that enable multiple folding pathways to converge toward the native state \cite{energy_landscape, Onuchic04COSB}. Furthermore, experimental studies have established native-state topology as an important determinant of protein folding rates \cite{baker_topology, hood2025exploring}, suggesting a close relationship between protein topology and foldability - an important determinant in natural selection. 

Protein foldability is closely related to collapsibility, which describes the propensity of a polypeptide chain to adopt compact conformations \cite{Thirumalai_1,Thirumalai_2,Thirumalai_3, sosnick2025collapse}. Although closely related, collapse and folding are distinct transitions: collapse involves chain compaction, whereas folding requires the formation of the specific native structure \cite{Thirumalai_1, kaushik2026intermediate}. Previous theoretical studies have shown that the separation between the sequence-dependent collapse and folding transition temperatures strongly correlates with folding rates \cite{Thirumalai_2,Thirumalai_3,Thirum95JPI}. When these transitions occur under similar thermodynamic conditions, proteins can fold rapidly without prolonged trapping in metastable intermediates \cite{Thirumalai_3}. Understanding the physical determinants of protein collapsibility is therefore essential for elucidating protein folding mechanisms.

Analytical models provide an efficient means of identifying these determinants across broad ranges of physical parameters. Previous work developed an analytical theory that characterizes protein collapsibility through the critical attractive interaction strength, $k_\theta$, required to overcome excluded-volume repulsion and induce chain collapse \cite{collapsibility,collapsibility_review}. A larger $k_\theta$ indicates a lower intrinsic propensity for collapse. This critical strength depends on both protein size, $N$, and native-state topology, suggesting that collapsibility may constrain evolutionary sequence selection and help explain preferences for particular protein topologies and multidomain architectures \cite{collapsibility,collapsibility_review}.

While protein size and native-state topology are key determinants of collapsibility, protein folding is also strongly influenced by their interaction with solvent\cite{fernandez2004solvation, solvent_role}, particularly water \cite{baron2012hydrophobic, clark2020water, best2020emerging, cheung2002protein}. During this process, Hydrostatic pressure can alter these interactions, promoting water penetration into protein cavities, structural rearrangements, and unfolding \cite{hillson1999pressure, pressure_stability,kauzmann,nagae2012high, royer2026mapping}. Pressure is also an important environmental factor in microbial evolution\cite{schimel2018life}, particularly in deep-sea and subsurface habitats that may have played a role in the origin of life \cite{origin1,origin2}. Natural selection may therefore favor protein structures and conformational dynamics adapted to various pressure. Understanding the contribution from pressure-dependent water interactions to protein collaspe may thus provide more thorough insights into the natural selection of protein structures. 

The original framework of collapsibility approximated residue–residue interactions using a Lennard-Jones potential with a single attractive minimum, which does not capture the complexity of water-mediated interactions. In aqueous environments, effective residue–residue interactions exhibit both direct-contact and water-mediated minima, separated by a desolvation barrier\cite{cheung2002protein, desolvation_baker} that influences protein folding kinetics. Incorporating a desolvation potential is therefore essential for describing how water interactions influence protein collapsibility. Previous studies have characterized the pressure dependence of water-mediated interactions and desolvation potentials \cite{hummer1998pressure,hillson1999pressure,Desolvation}. By altering the relative stability of direct-contact and water-mediated configurations, pressure provides a means of regulating protein conformational state. 

Here, we extend the analytical theory of protein collapsibility by incorporating pressure-dependent desolvation potentials into effective residue–residue interactions. We investigate how hydrostatic pressure influences collapsibility across proteins of different sizes and native-state topologies, focusing particularly on the role of non-local contacts. We further compare our theoretical predictions with experimental observations to elucidate the physical mechanisms underlying pressure-induced protein conformational transitions.

\section{Methods}

\subsection{Theoretical framework}

The collapsiblity theory \cite{collapsibility} employed a modified Edwards Hamiltonian for polymers \cite{Edward_Hamiltonian} to describe interactions within a polypeptide chain \cite{collapsibility}. In the continuum representation, the Hamiltonian is given by

\begin{equation} \label{eq1}
\mathcal{H}=\frac{3 k_{\mathrm{B}} T}{2 a_0^2} \int_0^N\left(\frac{\partial \mathbf{r}}{\partial s}\right)^2 \mathrm{~d} s+k_{\mathrm{B}} T \mathcal{V}(\mathbf{r}(s))
\end{equation}
where $\mathbf{r}(s)$ denotes the position of residue $s$, $N$ is the number of residues, $a_0$ is the diameter of an amino acid residue, and $k_{\mathrm{B}}T$ is the thermal energy. The first term describes chain connectivity, while the second term represents interactions between residues through the effective potential $\mathcal{V}(\mathbf{r}(s))$.

Following the theoretical framework of Samanta et al. \cite{collapsibility}, the interaction potential is modeled as a combination of a repulsive Gaussian, representing excluded-volume interactions, and an attractive Gaussian, representing interactions between native contact pairs:

\begin{equation} \label{eq2}
\begin{aligned}
\mathcal{V}(\mathbf{r}(s))= & \frac{\nu}{\left(2 \pi a_0^2\right)^{3 / 2}} \sum_{s=0}^N \sum_{s^{\prime}=0}^N \mathrm{e}^{-\frac{\left(\mathbf{r}(s)-\mathbf{r}(s^{\prime})\right)^2}{2 a_0^2}} \\
& -\frac{\kappa(p)}{\left(2 \pi \sigma(p)^2\right)^{3 / 2}} \sum_{\left\{s_i, s_j\right\}} \mathrm{e}^{-\frac{\left(\mathbf{r}(s_i)-\mathbf{r}(s_j)\right)^2}{2 \sigma(p)^2}}
\end{aligned}
\end{equation}
where $\nu$ is the excluded-volume interaction coefficient, set to $\frac{4}{3}\pi a_0^3$, and $\kappa(p)$ is the attractive interaction coefficient. It is worth noting that the potential of mean force between amino acids (charged, polar, and hydrophobic residues) have direct contacts and other solvent separated minima~\cite{Vaitheeswaran08PNAS}.  Thus, Eq. \ref{eq2} should be thought of a simplified but generic representation of the interactions between amino acid residues. The parameter $\sigma(p)$ in Eq. \ref{eq2} characterizes the spatial range of attractive interactions between native contact pairs. We note parenthetically that the sum of two Gaussian  terms is a reasonable mimic of the Lennard-Jones interaction~\cite{Singer58Nature} with the advantage that the calculations can be done analytically.  In the original framework, the interaction potential consists of a single effective attractive well without explicitly accounting for pressure-dependent desolvation.

To assess the pressure dependence on collapsibilty, we extended this framework using the desolvation potential described elsewhere \cite{Desolvation} (see Appendix B). As illustrated in Figure~\ref{fig:method}, the desolvation potential contains two local minima. The first minimum corresponds to direct residue-residue contact (contact minimum), while the second minimum represents a configuration in which a water molecule separates the two residues (water-mediated minimum).

\begin{figure*}[htbp]
    \centering
    \includegraphics[width=\textwidth]{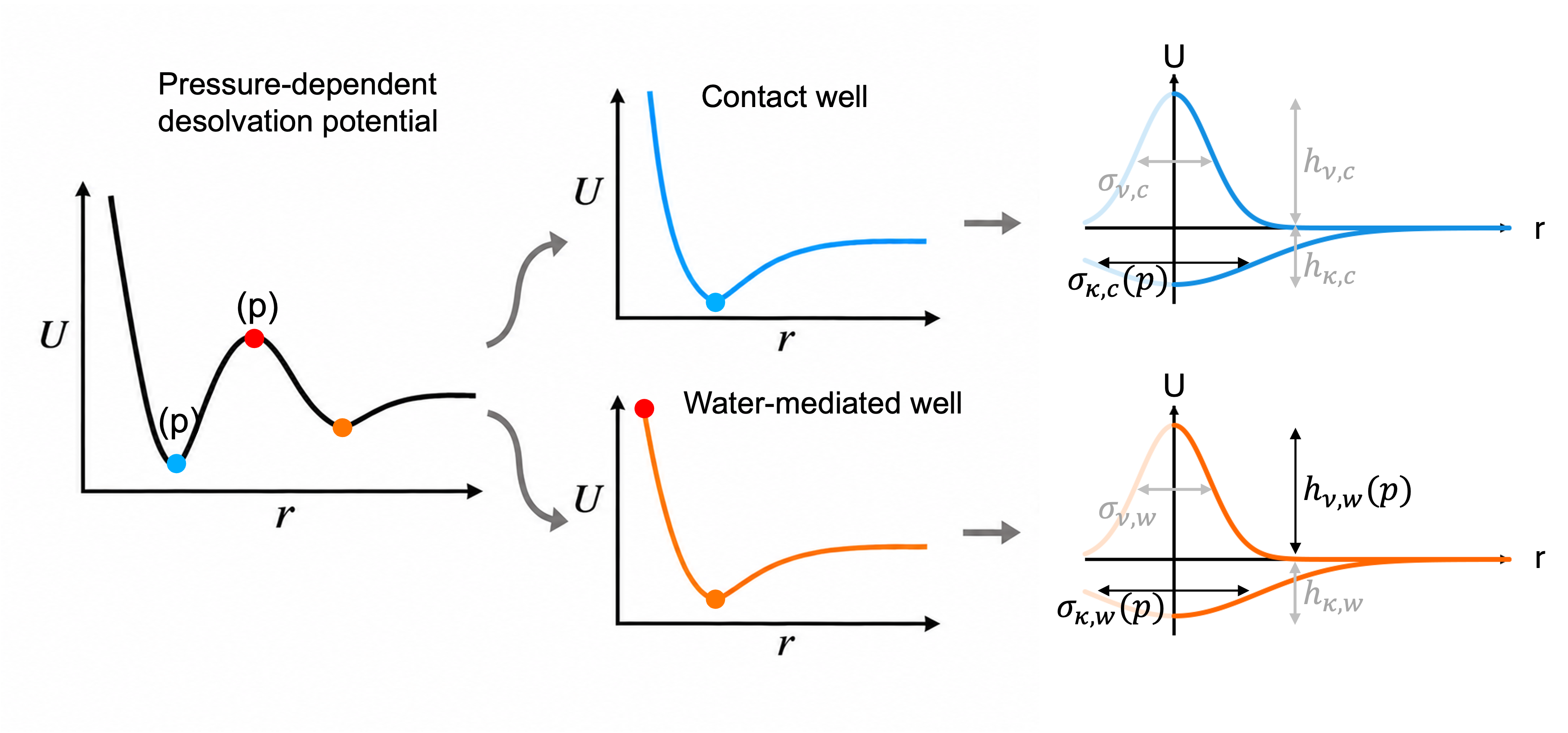}
    \caption{
    \textbf{Schematic illustration of the Gaussian approximation of the pressure-dependent desolvation potential.}
    The original desolvation potential (left), consisting of a contact well and a water-mediated well separated by a desolvation barrier, is approximated by two separate single-well potentials (middle).
    The blue and orange dots indicate the minima of the contact and water-mediated wells, respectively, while the red dots indicate the desolvation barrier and its corresponding position in the approximated water-mediated potential. Pressure-dependent positions in the curve are indicated by $(p)$. 
    Each single-well potential is further represented as the sum of a repulsive and an attractive Gaussian (right), characterized by their amplitudes $h_{\nu}$ and $h_{\kappa}$ and widths $\sigma_{\nu}$ and $\sigma_{\kappa}$, respectively.
    Subscripts $c$ and $w$ denote the contact and water-mediated wells, with pressure-dependent parameters shown in black and pressure-independent parameters in gray.
    }
    \label{fig:method}
\end{figure*}

To investigate the contributions of these two interaction states to protein collapsibility, we approximated the contact and water-mediated wells separately. Each well was represented by the sum of a repulsive Gaussian and an attractive Gaussian, consistent with the functional form used in the original collapsibility theory~\cite{collapsibility}. This decomposition allows us to characterize the pressure dependence of the two interaction states independently while retaining the Gaussian representation required for the analytical calculation of collapsibility.

We considered pressures ranging from $p=0$ to $p=5$, where $p=0$ corresponds to atmospheric pressure and one pressure unit corresponds to approximately 760 bar \cite{Desolvation}. For both wells, pressure dependence was introduced through changes in the attractive Gaussian width, while the repulsive Gaussian width remained constant. This choice was made because we think the repulsion Gaussian width should be a representation of residue size, which stays constant in the model. The pressure-dependent attractive Gaussian widths were determined numerically by matching the effective Gaussian potentials to the pressure-dependent desolvation energies, as described below.

\subsection{Gaussian approximation of the desolvation potential}

Each of the two potential wells was approximated by the combination of a repulsive and an attractive Gaussian:

\begin{equation}
U(r)=h_\nu \mathrm{e}^{-\frac{r^2}{2\sigma_\nu^2}}-h_\kappa \mathrm{e}^{-\frac{r^2}{2\sigma_\kappa^2}}
\label{Eq:Gaussian}
\end{equation}
where $r$ is the distance between two residues, $h_\nu$ and $h_\kappa$ are the amplitudes of the repulsive and attractive Gaussians, respectively, and $\sigma_\nu$ and $\sigma_\kappa$ are their corresponding widths. The negative sign of the attractive term represents its favorable contribution to the interaction energy. The combination of these two Gaussians produces a potential well with a repulsive region at short distances and an attractive minimum at larger distances.
The  parameters in Eq. \ref{Eq:Gaussian} were determined separately for the contact and water-mediated wells. Their pressure-dependent values are summarized in Table I and Table II, respectively.

\subsubsection{Determination of Gaussian widths at $p=0$}

For the contact well, the repulsive and attractive Gaussian widths at $p=0$ were adopted from the original collapsibility framework~\cite{collapsibility}, where $p=0$ corresponds to atmospheric pressure and one pressure unit corresponds to approximately 760 bar \cite{Desolvation}. The resulting widths are listed in Table I.

For the water-mediated well, the repulsive Gaussian width was kept identical to that of the contact well. The attractive Gaussian width was increased by the diameter of a water molecule (0.28 nm) to account for the additional separation between residues introduced by an intervening water molecule. The resulting widths are listed in Table II.

\subsubsection{Determination of Gaussian amplitudes at $p=0$}

The Gaussian amplitudes were selected to approximate the shape of the desolvation potential while reproducing the corresponding well depths and barrier height at atmospheric pressure.

For the contact well, the repulsive and attractive Gaussian amplitudes were chosen to reproduce the depth of the contact minimum while approximating the slope of the repulsive region of the original desolvation potential. The resulting amplitudes are listed in Table I.

For the water-mediated well, the amplitudes were constrained such that their combined contribution at the reference separation reproduced the desolvation barrier height. The individual amplitudes were then selected to reproduce the water-mediated well depth while approximating the slope of its repulsive region. The resulting amplitudes are listed in Table II.

\subsubsection{Pressure dependence of the Gaussian potentials}

To incorporate pressure dependence, we used the pressure-dependent well depths and desolvation barrier described in Appendix B. As pressure increases, the contact well becomes shallower, while the desolvation barrier increases. The water-mediated well depth remains constant within the pressure range considered\cite{Desolvation}.

For the contact well, both Gaussian amplitudes and the repulsive Gaussian width were held constant, while the attractive Gaussian width was varied to reproduce the pressure-dependent contact-well depth. The resulting pressure-dependent Gaussian parameters are summarized in Table I.

For the water-mediated well, the attractive Gaussian amplitude and repulsive Gaussian width were held constant. The repulsive Gaussian amplitude was allowed to increase with pressure to reproduce the increasing desolvation barrier, while the attractive Gaussian width was adjusted to maintain the prescribed water-mediated well depth. The resulting pressure-dependent Gaussian parameters are summarized in Table II.

\subsection{Calculation of pressure-dependent collapsibility}

\subsubsection{Determination of pressure-dependent Gaussian widths}

To determine the pressure-dependent attractive Gaussian widths, we first derived an analytical expression for the well depth of the two-Gaussian potential. By solving for the position of the local minimum, $r_{min}$, and evaluating the potential at this position, the minimum energy can be expressed as

\begin{equation}
U(r_{min})=-h_\kappa\left(\frac{h_\kappa\sigma_\nu^2}{h_\nu\sigma_\kappa^2}\right)^{-\frac{\sigma_\nu^2}{\sigma_\nu^2-\sigma_\kappa^2}}+h_\nu\left(\frac{h_\kappa\sigma_\nu^2}{h_\nu\sigma_\kappa^2}\right)^{-\frac{\sigma_\kappa^2}{\sigma_\nu^2-\sigma_\kappa^2}}
\end{equation}
where $h_\nu$ and $h_\kappa$ denote the repulsive and attractive Gaussian amplitudes, respectively, and $\sigma_\nu$ and $\sigma_\kappa$ are their corresponding widths. This expression was applied independently to the contact and water-mediated wells.

The pressure dependence of the well depths and desolvation barrier was obtained from Equation B1 in Appendix B. By substituting these relationships into the analytical expression for the minimum energy, we numerically solved for the attractive Gaussian width $\sigma_\kappa(p)$ at each pressure.

For the contact well, the Gaussian amplitudes and repulsive Gaussian width were fixed at their atmospheric-pressure values (Table I). The pressure-dependent attractive Gaussian width $\sigma_{\kappa,c}(p)$ was determined by solving
    
\begin{equation}
\begin{aligned}
-h_{\kappa, c}\left(\frac{h_{\kappa, c}\sigma_{\nu, c}^2}{h_{\nu, c}\sigma_{\kappa, c}^2}\right)^{-\frac{\sigma_{\nu, c}^2}{\sigma_{\nu, c}^2-\sigma_{\kappa, c}^2}}+ &h_{\nu, c}\left(\frac{h_{\kappa, c}\sigma_{\nu, c}^2}{h_{\nu, c}\sigma_{\kappa, c}^2}\right)^{-\frac{\sigma_{\kappa, c}^2}{\sigma_{\nu, c}^2-\sigma_{\kappa, c}^2}}\\
&= \epsilon_0-0.076p
\end{aligned}
\end{equation}
where the subscript $c$ denotes the contact well, and $\epsilon_0=0.6$ specifies the reference contact-well depth (absolute value) at atmospheric pressure.

For the water-mediated well, the attractive Gaussian amplitude and repulsive Gaussian width were fixed at their atmospheric-pressure values (Table II). The repulsive Gaussian amplitude $h_{\nu,w}(p)$ and attractive Gaussian width $\sigma_{\kappa,w}(p)$ were determined simultaneously by imposing two constraints: the water-mediated well depth remains constant, while the desolvation barrier increases with pressure.

The corresponding equations are,

\begin{equation}
\begin{aligned}
-h_{\kappa, w}\left(\frac{h_{\kappa, w}\sigma_{\nu, w}^2}{h_{\nu, w}\sigma_{\kappa, w}^2}\right)^{-\frac{\sigma_{\nu, w}^2}{\sigma_{\nu, w}^2-\sigma_{\kappa, w}^2}}&+h_{\nu, w}\left(\frac{h_{\kappa, w}\sigma_{\nu, w}^2}{h_{\nu, w}\sigma_{\kappa, w}^2}\right)^{-\frac{\sigma_{\kappa, w}^2}{\sigma_{\nu, w}^2-\sigma_{\kappa, w}^2}}\\
&= \epsilon_0'\\
\end{aligned}
\end{equation}

\begin{equation}
h_{\nu, w}-h_{\kappa, w} = \epsilon_0'' + 0.127p
\end{equation}
where the subscript $w$ denotes the water-mediated well, $\epsilon_0'=0.2$ is the reference water-mediated well depth (absolute value), and $\epsilon_0''=0.8$ is the reference desolvation barrier height at atmospheric pressure.

By numerically solving these equations at each pressure, we obtained the pressure-dependent attractive Gaussian widths for the contact and water-mediated wells.

\subsubsection{Calculation of the critical attraction strength}

Using the pressure-dependent attractive Gaussian widths determined above, we calculated the collapsibility (critical attraction strength $\kappa_\theta(p)$ )for each well. Following the theoretical framework, derived in Appendix A (Equation A3),  the critical strength of attraction is given by,

\begin{equation}
\begin{aligned}
\kappa_\theta(p)=\frac{4}{3} \pi a_0^3 \frac{\sum_{s=0}^N \sum_{s^{\prime}=0}^N \frac{\sum_{n=1}^N \frac{1-\cos \left[n \pi\left(s-s^{\prime}\right) / N\right]}{n^4}}{\left(\sum_{n=1}^N \frac{1-\cos \left[n \pi\left(s-s^{\prime}\right) / N\right]}{n^2}+\frac{3 \pi^2}{2 N}\right)^{5 / 2}}}{\sum_{\left\{s_i, s_j\right\}} \frac{\sum_{n=1}^N \frac{1-\cos \left[n \pi\left(s_i-s_j\right) / N\right]}{n^4}}{\left(\sum_{n=1}^N \frac{1-\cos \left[n \pi\left(s_i-s_j\right) / N\right]}{n^2}+\frac{3 \pi^2 \sigma_\kappa(p)^2}{2 a_0^2 N}\right)^{5 / 2}}}
\end{aligned}
\end{equation}
where $\sigma_\kappa(p)$ is the pressure-dependent attractive Gaussian width obtained from the corresponding effective potential. All other parameters in this expression are independent of pressure.

The calculation was performed separately for the contact and water-mediated wells, allowing us to quantify their individual contributions to pressure-dependent protein collapsibility. In this theory, $\kappa_\theta(p)$ represents the critical attraction required for polymer collapse, providing a quantitative measure of the tendency of a protein sequence to undergo a coil-to-globule transition under different pressures.

\section{Results}

\subsection{Pressure-dependent changes in the contact and water-mediated potentials}

The Gaussian parameters obtained for the contact and water-mediated potentials at different pressures are summarized in Tables I and II, respectively.

\begin{table}
\begin{tabular}{|p{1.5cm}|p{1.5cm}|p{1.5cm}|p{1.5cm}|p{1.5cm}|}
\hline
$p$ & $\sigma_{\kappa, c}$ & $\sigma_{\nu, c}$ & $h_{\kappa, c}$ & $h_{\nu, c}$ \\
\hline
0 & 0.63 & 0.38 & 5 & 33.62 \\
\hline
1 & 0.61 & 0.38 & 5 & 33.62 \\
\hline
2 & 0.60 & 0.38 & 5 & 33.62 \\
\hline
3 & 0.58 & 0.38 & 5 & 33.62 \\
\hline
4 & 0.57 & 0.38 & 5 & 33.62 \\
\hline
5 & 0.55 & 0.38 & 5 & 33.62 \\
\hline
\end{tabular}
\caption{\label{tab:contact_parameters} Pressure-dependent Gaussian parameters for the contact well. $\sigma_{\kappa,c}$ and $\sigma_{\nu,c}$ denote the attractive and repulsive Gaussian widths, respectively, while $h_{\kappa,c}$ and $h_{\nu,c}$ denote their corresponding amplitudes. Gaussian widths are expressed in nm. Pressure follows the convention of Gasic et al. \cite{Desolvation}, where $p=0$ represents atmospheric pressure and one pressure unit corresponds to approximately 760 bar.}
\end{table}

\begin{table}
\begin{tabular}{ |m{1.5cm}|m{1.5cm}|m{1.5cm}|m{1.5cm}|m{1.5cm}|}
\hline
$p$ & $\sigma_{\kappa, w}$ & $\sigma_{\nu, w}$ & $h_{\kappa, w}$ & $h_{\nu, w}$ \\
\hline
0 & 0.92 & 0.38 & 0.43 & 1.23 \\
\hline
1 & 0.93 & 0.38 & 0.43 & 1.36 \\
\hline
2 & 0.95 & 0.38 & 0.43 & 1.48 \\
\hline
3 & 0.96 & 0.38 & 0.43 & 1.61 \\
\hline
4 & 0.97 & 0.38 & 0.43 & 1.75 \\
\hline
5 & 0.98 & 0.38 & 0.43 & 1.87 \\
\hline
\end{tabular}
\caption{\label{tab:water_parameters} Pressure-dependent Gaussian parameters for the water-mediated well. $\sigma_{\kappa,w}$ and $\sigma_{\nu,w}$ denote the attractive and repulsive Gaussian widths, respectively, while $h_{\kappa,w}$ and $h_{\nu,w}$ denote their corresponding amplitudes. Gaussian widths are expressed in nm. Pressure follows the convention of Gasic et al. \cite{Desolvation}, where $p=0$ represents atmospheric pressure and one pressure unit corresponds to approximately 760 bar.}
\end{table}

\begin{figure*}
    \centering
    \begin{subfigure}{0.49\textwidth}
    \includegraphics[width=\textwidth]{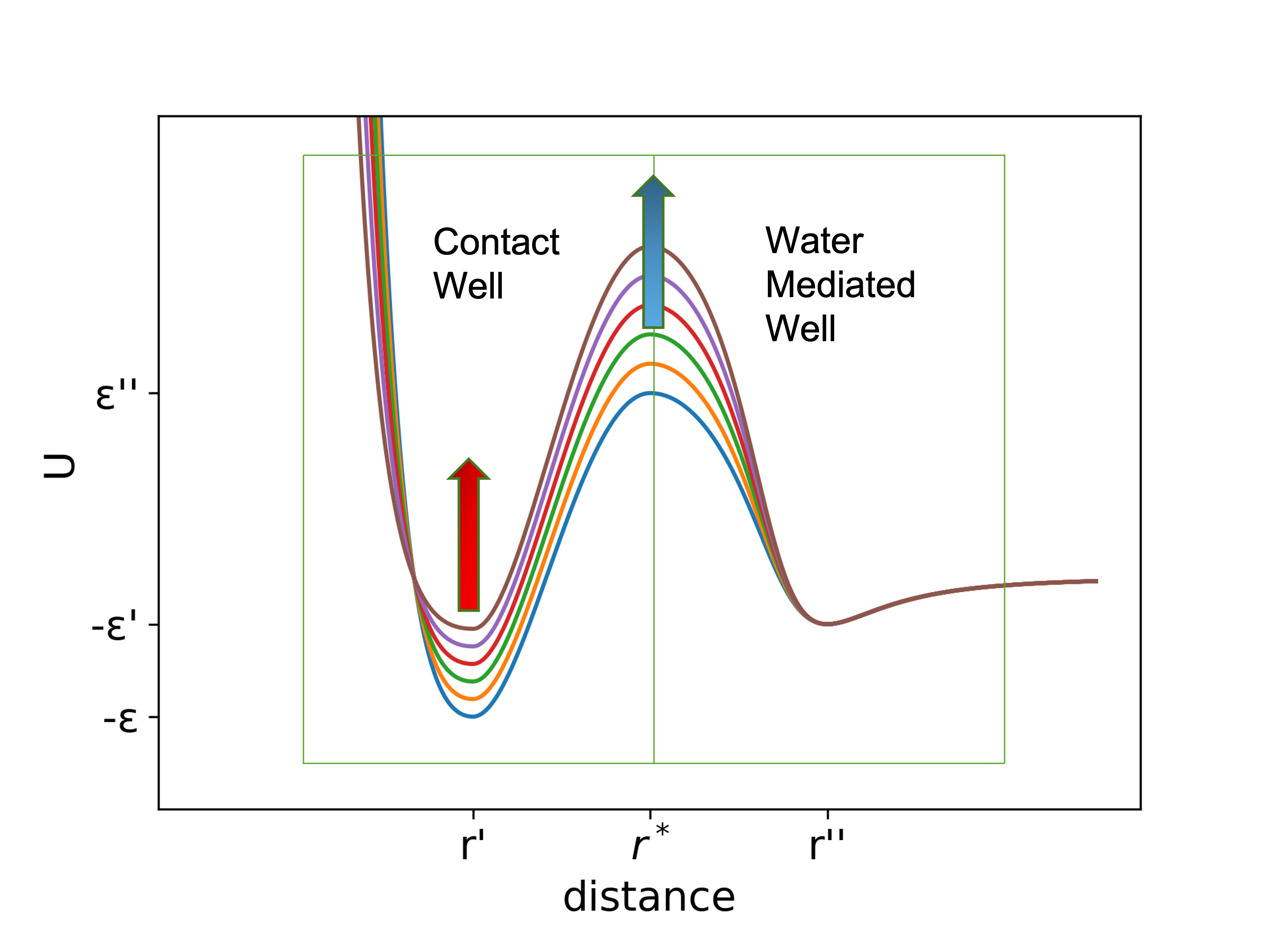}
    \caption{Original desolvation potential}
    \end{subfigure}
    \hfill
    \begin{subfigure}{0.49\textwidth}
    \includegraphics[width=\textwidth]{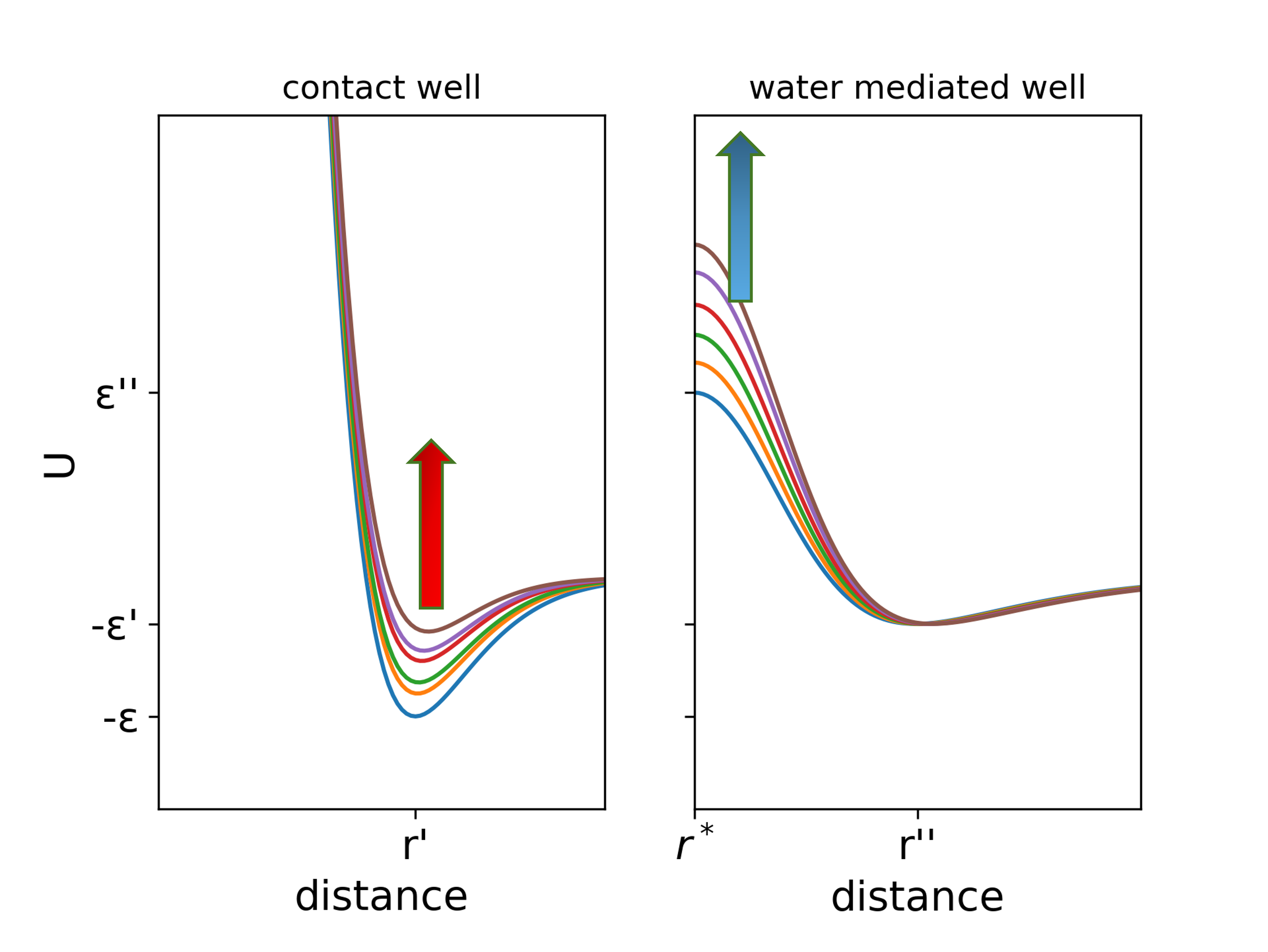}
    \caption{Approximated separated wells}
    \end{subfigure}
\caption{\label{fig:potential} Comparison of (a) the original pressure-dependent desolvation potential developed by Gasic et al. \cite{Desolvation} and (b) the separately approximated contact and water-mediated wells using the Gaussian parameters in Tables I and II. Curves represent pressures from $p=0$ (blue) to $p=5$ (brown), with arrows indicating increasing pressure. The contact well describes direct residue-residue interactions, while the water-mediated well represents interactions involving an intervening water molecule. The starting point of the water-mediated well corresponds to the desolvation barrier position $r^*$. One pressure unit corresponds to approximately 760 bar.}
\end{figure*}

The resulting pressure-dependent Gaussian potentials are shown in Figure~\ref{fig:potential}, alongside the original desolvation potential developed by Gasic et al. \cite{Desolvation}. As pressure increases, the contact well becomes progressively shallower, reflecting the destabilization of direct residue-residue contacts. In contrast, the water-mediated well maintains approximately the same minimum energy while the repulsive barrier increases. These trends reproduce the distinct pressure responses of the contact and water-mediated interaction states in the original desolvation model.

To investigate the pressure dependence of protein collapsibility, we calculated the collapsibility associated with the contact and water-mediated wells, denoted by $\kappa_{\theta,c}$ and $\kappa_{\theta,w}$, respectively, for 30 proteins over the pressure range $p=0$ to $p=5$. Following the definition in Samanta et al. \cite{collapsibility}, $k_\theta = (2\pi\sigma^2)^{-3/2}\kappa_\theta$ is the average attractive interaction strength (in units of $k_BT$) of a contact at the critical point. A larger value of $k_\theta$ indicates lower collapsibility, whereas a smaller value indicates greater collapsibility.

We further defined the water-mediated collapsibility ratio $\eta$ as
\begin{equation}
\eta = \frac{k_{\theta,w}}{k_{\theta,c}+k_{\theta,w}},
\end{equation}
which represents the relative contribution of water-mediated collapsibility compared with the combined collapsibility of the native-contact and water-mediated states. A smaller value of $\eta$ indicates a smaller relative collapsibility parameter associated with the water-mediated state. Together, $k_{\theta,c}$, $k_{\theta,w}$, and $\eta$ are referred to as collapsibility metrics.

\subsection{Increasing pressure reduces native-state collapsibility while favoring water-mediated collapse}

\begin{figure}
    \centering
    \includegraphics[width=9cm]{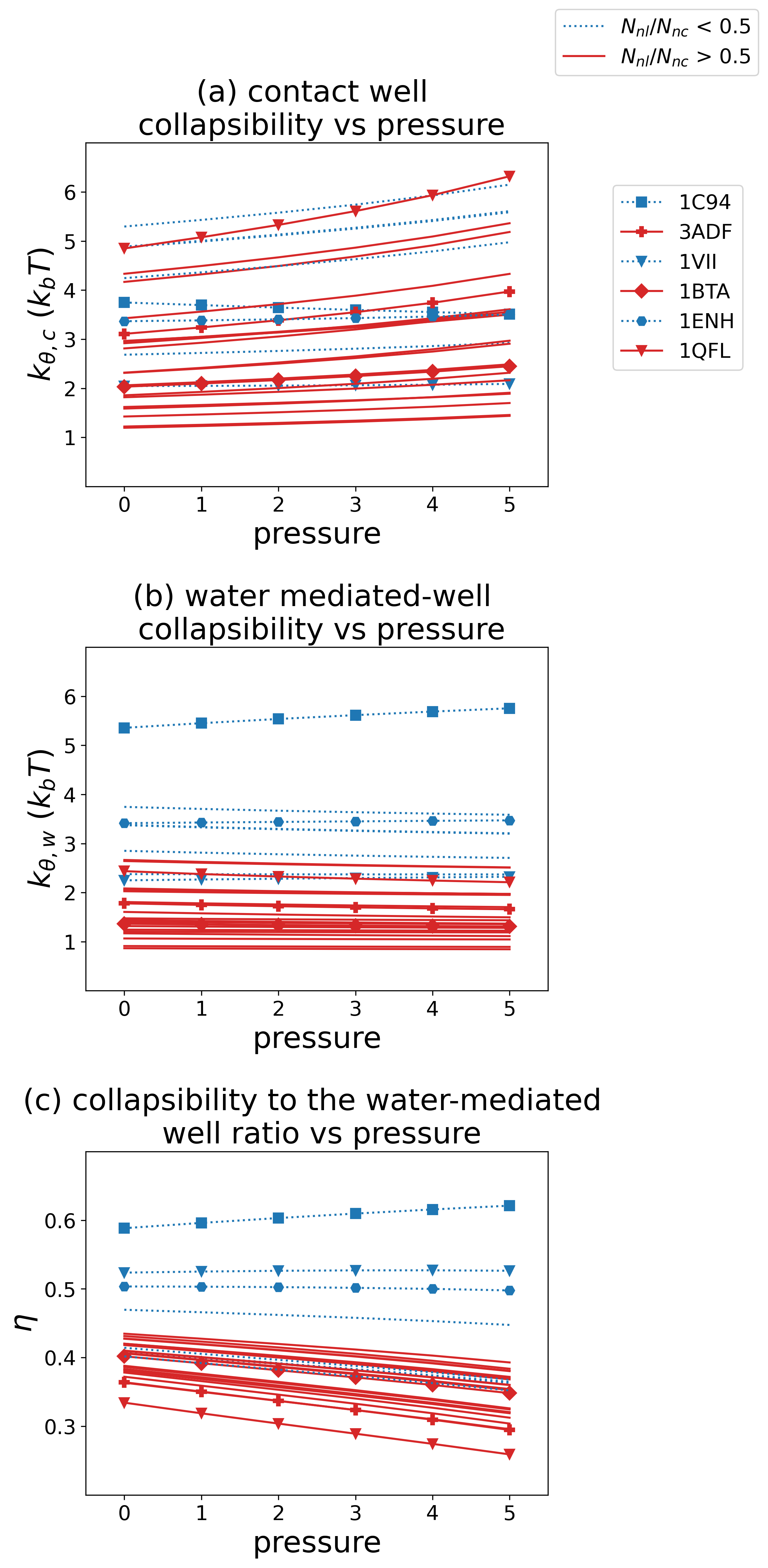}
    \caption{\label{fig:collapsibility_pressure} Pressure dependence of (a) contact-well collapsibility $k_{\theta,c}$, (b) water-mediated-well collapsibility $k_{\theta,w}$, and (c) water-mediated collapsibility ratio $\eta$ for 30 proteins over $p=0$ to $p=5$. Proteins are classified as $\alpha$-rich or $\beta$-rich according to their non-local native contact ratio $N_{nl}/N_{nc}$ \cite{collapsibility}, where $N_{nl}$ denotes the number of non-local native contacts and $N_{nc}$ denotes the total number of native contacts. Markers identify selected proteins examined in greater detail in Figure~\ref{fig:structures}. One pressure unit corresponds to approximately 760 bar.}
\end{figure}

The pressure dependence of the three collapsibility metrics is shown in Figure~\ref{fig:collapsibility_pressure}. Proteins were classified according to their native-state non-local contact ratio, $N_{nl}/N_{nc}$, following Samanta et al. \cite{collapsibility}, where $N_{nl}$ denotes the number of non-local native contacts and $N_{nc}$ denotes the total number of native contacts. Proteins with $N_{nl}/N_{nc}<0.5$ were classified as $\alpha$-rich, whereas proteins with $N_{nl}/N_{nc}>0.5$ were classified as $\beta$-rich.

As shown in Figure~\ref{fig:collapsibility_pressure}(a), contact well collapsibility $k_{\theta,c}$ increases consistently with pressure, indicating that proteins become less collapsible toward the native-contact state. This trend is consistent with the destabilization of direct residue-residue contacts during pressure-induced denaturation.

In contrast, Figure~\ref{fig:collapsibility_pressure}(b) shows that water-mediated well collapsibility $k_{\theta,w}$ decreases slightly for most proteins as pressure increases, indicating that these sequences become more collapsible toward the water-mediated state. The water-mediated collapsibility ratio $\eta$ also decreases substantially with pressure (Figure~\ref{fig:collapsibility_pressure}(c)), indicating that the collapsibility parameter associated with the water-mediated state becomes smaller relative to the combined collapsibility parameters of the two states.

Together, these results suggest that pressure-induced denaturation does not necessarily eliminate the tendency of proteins to form compact conformations. Although increasing pressure reduces native-state collapsibility, proteins can retain or even exhibit enhanced collapsibility toward water-mediated configurations. This distinction suggests that disruption of native contacts under pressure may coexist with the persistence of partially collapsed, water-mediated conformations.

\subsection{Larger proteins and those with more non-local native contacts exhibit greater pressure sensitivity and enhanced water-mediated collapsibility}
The pressure dependence of protein collapsibility is strongly influenced by native contact topology. As shown in Figures~\ref{fig:collapsibility_pressure}(a) and \ref{fig:collapsibility_pressure}(c), $\beta$-rich proteins, which generally contain a higher proportion of non-local native contacts, exhibit greater pressure-induced changes in collapsibility than $\alpha$-rich proteins. To quantify this difference, we calculated the changes in $k_{\theta,c}$, $k_{\theta,w}$, and $\eta$ between $p=0$ and $p=5$ for each protein and averaged them within the two structural categories (Table~\ref{tab:collapsibility_changes}).

\begin{table}
\begin{tabular}{ |m{1.5cm}|m{1.5cm}|m{1.5cm}|m{1.5cm}|}
\hline
 & $\Delta k_{\theta,c}$ & $\Delta k_{\theta,w}$ & $\Delta\eta$ \\
\hline
$\alpha$-rich & 0.40 & -0.02 & -0.02 \\
\hline
$\beta$-rich & 0.59 & -0.08 & -0.06 \\
\hline
\end{tabular}
\caption{\label{tab:collapsibility_changes} Average changes in contact-well collapsibility $\Delta k_{\theta,c}$, water-mediated-well collapsibility $\Delta k_{\theta,w}$, and water-mediated collapsibility ratio $\Delta\eta$ from $p=0$ to $p=5$ for $\alpha$-rich and $\beta$-rich proteins. Changes are defined as the value at $p=5$ minus the value at $p=0$.}
\end{table}

On average, $\beta$-rich proteins exhibit a larger increase in $k_{\theta,c}$ than $\alpha$-rich proteins (0.59 versus 0.40), indicating a greater pressure-induced reduction in contact-well collapsibility. More notably, the decrease in $k_{\theta,w}$ is approximately four times greater in $\beta$-rich proteins (-0.08 versus -0.02), indicating a substantially enhanced propensity to collapse into water-mediated configurations under elevated pressure. The magnitude of the decrease in $\eta$ is also approximately three times greater (-0.06 versus -0.02), consistent with a stronger pressure-induced shift toward water-mediated collapsed states. Together, these results suggest that proteins with more non-local native contacts are not only more sensitive to pressure but also exhibit a greater enhancement of water-mediated collapsibility.

\begin{figure}
\begin{subfigure}{0.55\textwidth}
\includegraphics[width=\textwidth]{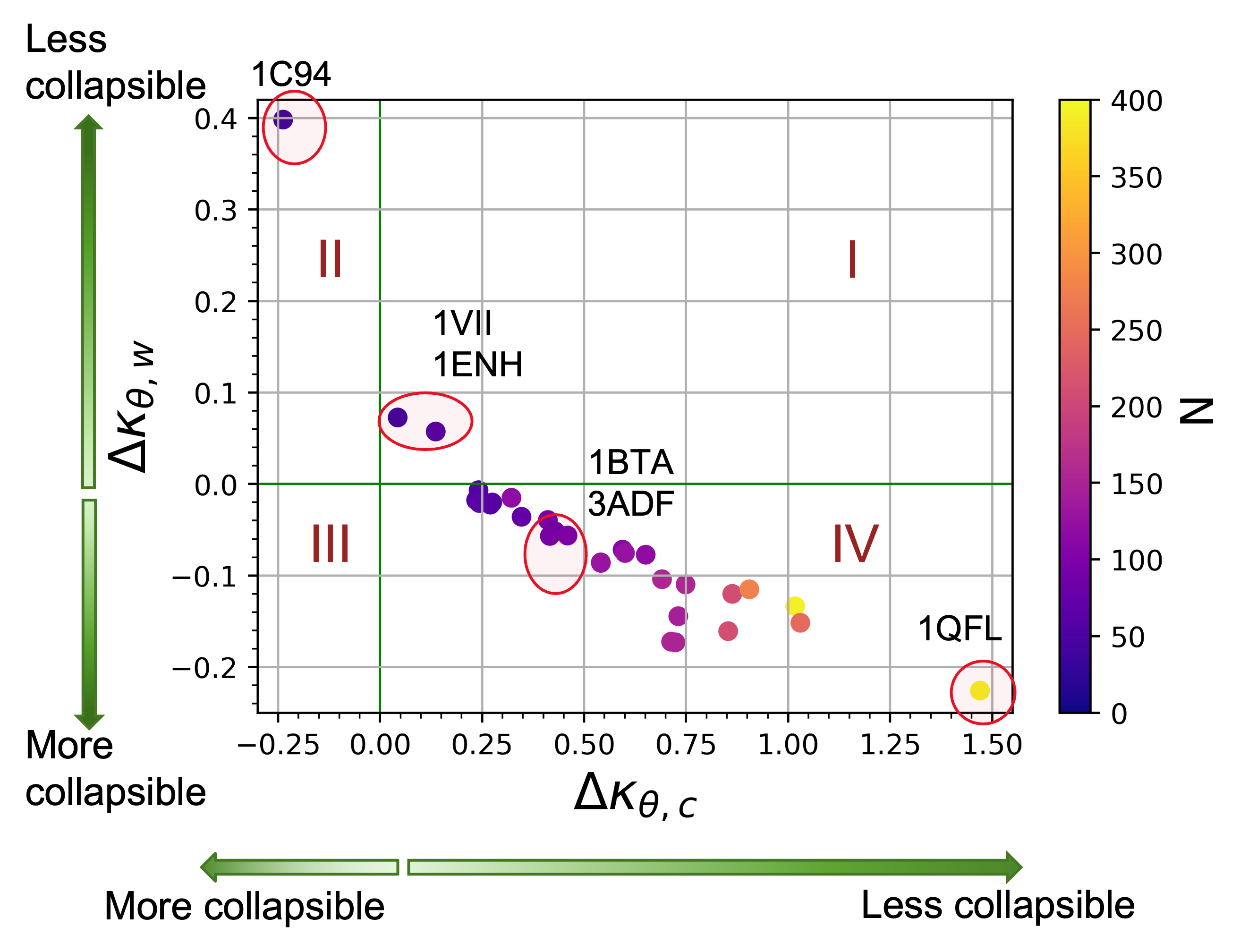}
\caption{}
\end{subfigure}
\hfill
\begin{subfigure}{0.4\textwidth}
\includegraphics[width=\textwidth]{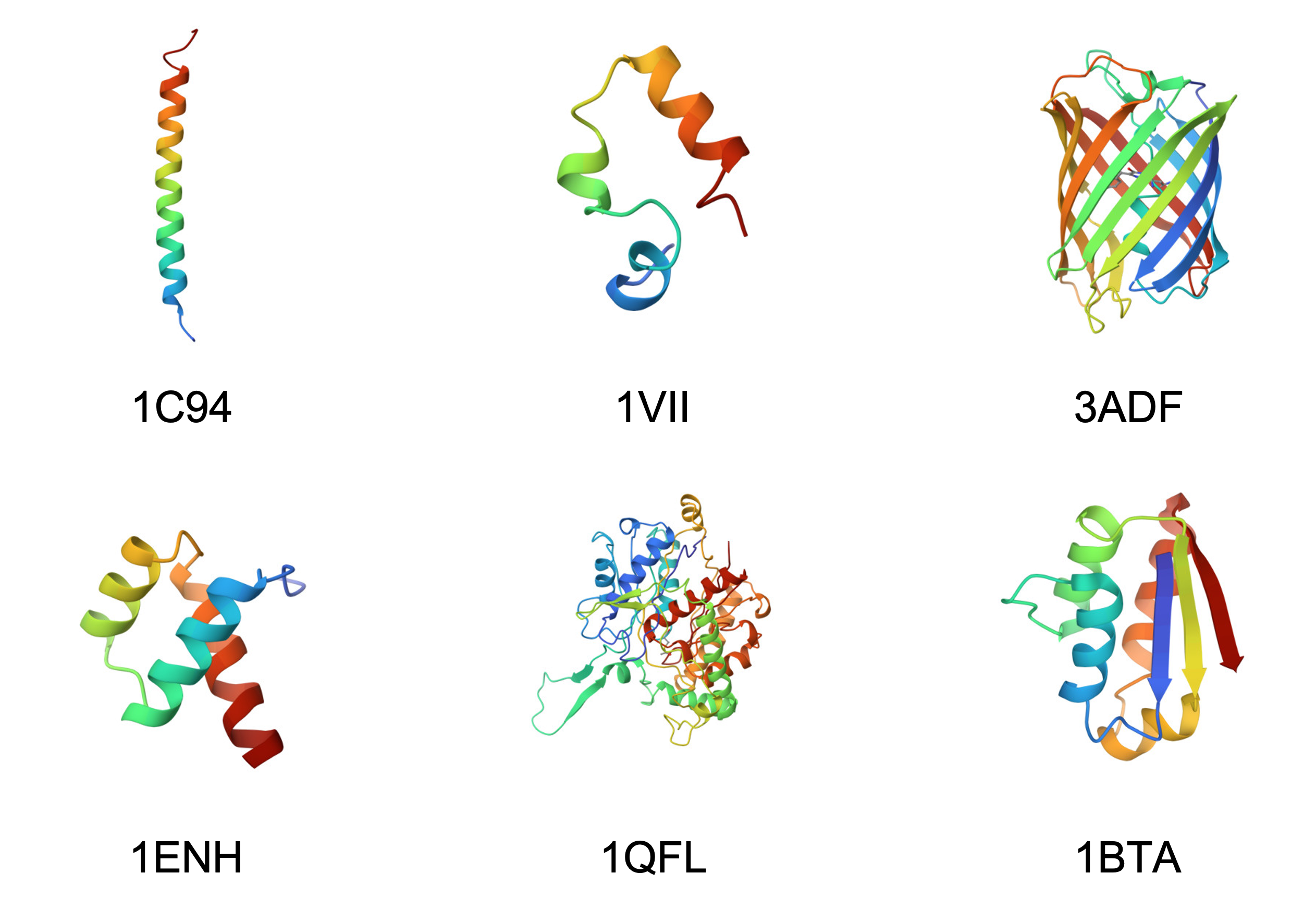}
\caption{}
\end{subfigure}
\caption{\label{fig:structures} (a) Changes in contact-well collapsibility $\Delta k_{\theta,c}$ (x-axis) and water-mediated-well collapsibility $\Delta k_{\theta,w}$ (y-axis) from $p=0$ to $p=5$ for 30 proteins (PDB IDs: 1MBO, 1IV9, 1MBN, 1GB0, 1C94, 1L0Q, 1BEB, 1M8N, 2LYZ, 1BDD, 1TFP, 3ADF, 2RN2, 1VII, 5O65, 1IGC, 2ABK, 1X3D, 1UBQ, 1R2R, 3CHY, 1BTA, 1A3N, 1SRL, 1NEB, 1K53, 1GM1, 1G6P, 1ENH, 1QFL). Color indicates the number of residues $N$. Positive changes indicate reduced collapsibility, whereas negative changes indicate enhanced collapsibility. The plane is divided into four quadrants according to the signs of the two changes. Six representative proteins are labeled by their PDB IDs. (b) Structures of the six representative proteins highlighted in panel (a).}
\end{figure}

To investigate protein-specific variations, we plotted $\Delta k_{\theta,c}$ against $\Delta k_{\theta,w}$ for all 30 proteins (Figure~\ref{fig:structures}(a)). Positive changes indicate reduced collapsibility, whereas negative changes indicate enhanced collapsibility. Each point is colored by protein size, $N$, and six representative structures are shown in Figure~\ref{fig:structures}(b).

Most proteins fall within the fourth quadrant ($\Delta k_{\theta,c}>0$, $\Delta k_{\theta,w}<0$), indicating that elevated pressure suppresses collapse into direct-contact configurations while promoting water-mediated collapse. This behavior is observed across different secondary-structure compositions, including predominantly $\alpha$-helical (1QFL), mixed $\alpha$/$\beta$ (1BTA), and predominantly $\beta$-sheet (3ADF) proteins. Thus, although the magnitude of the pressure response depends on native contact topology, the tendency toward water-mediated collapse is shared by proteins with diverse structures. A few proteins deviate from this general trend. The retro-GCN4 leucine zipper (1C94), consisting predominantly of a single $\alpha$ helix, falls within the second quadrant, indicating enhanced contact-well collapsibility under pressure. In contrast, two small three-helix proteins, 1ENH and 1VII, fall within the first quadrant, exhibiting reduced collapsibility toward both contact and water-mediated states. These proteins may therefore be more susceptible to pressure-induced expansion or the formation of partially unfolded conformations.

Protein size also contributes to the observed variation. Larger proteins generally exhibit greater pressure-induced changes, particularly larger decreases in $k_{\theta,w}$. Overall, these results suggest that both chain length and native contact topology influence pressure sensitivity, with proteins that are larger and containing more non-local contacts showing a particularly strong tendency toward water-mediated collapse under elevated pressure.

\subsection{Secondary-structure organization influences regional collapsibility under pressure}

The preceding results suggest that native contact topology contributes to the pressure dependence of protein collapsibility. To further investigate how these effects arise from different structural regions, we calculated $k_{\theta,c}$, $k_{\theta,w}$, and $\eta$ for selected regions of four representative proteins: the B domain of Staphylococcus aureus protein A (1BDD), single-chain monellin (1IV9), wild-type barstar (1BTA), and 3-phosphoglycerate kinase (1QPG).

These proteins represent different structural organizations, including three-helix bundles, mixed $\alpha$/$\beta$ structures, and multidomain proteins. The regional collapsibility metrics were compared with those calculated for the corresponding full-length sequences to assess the contributions of local secondary structures and longer-range interactions.

\begin{figure*}
\centering
\captionsetup[subfigure]{labelformat=empty}
\includegraphics[width=\textwidth]{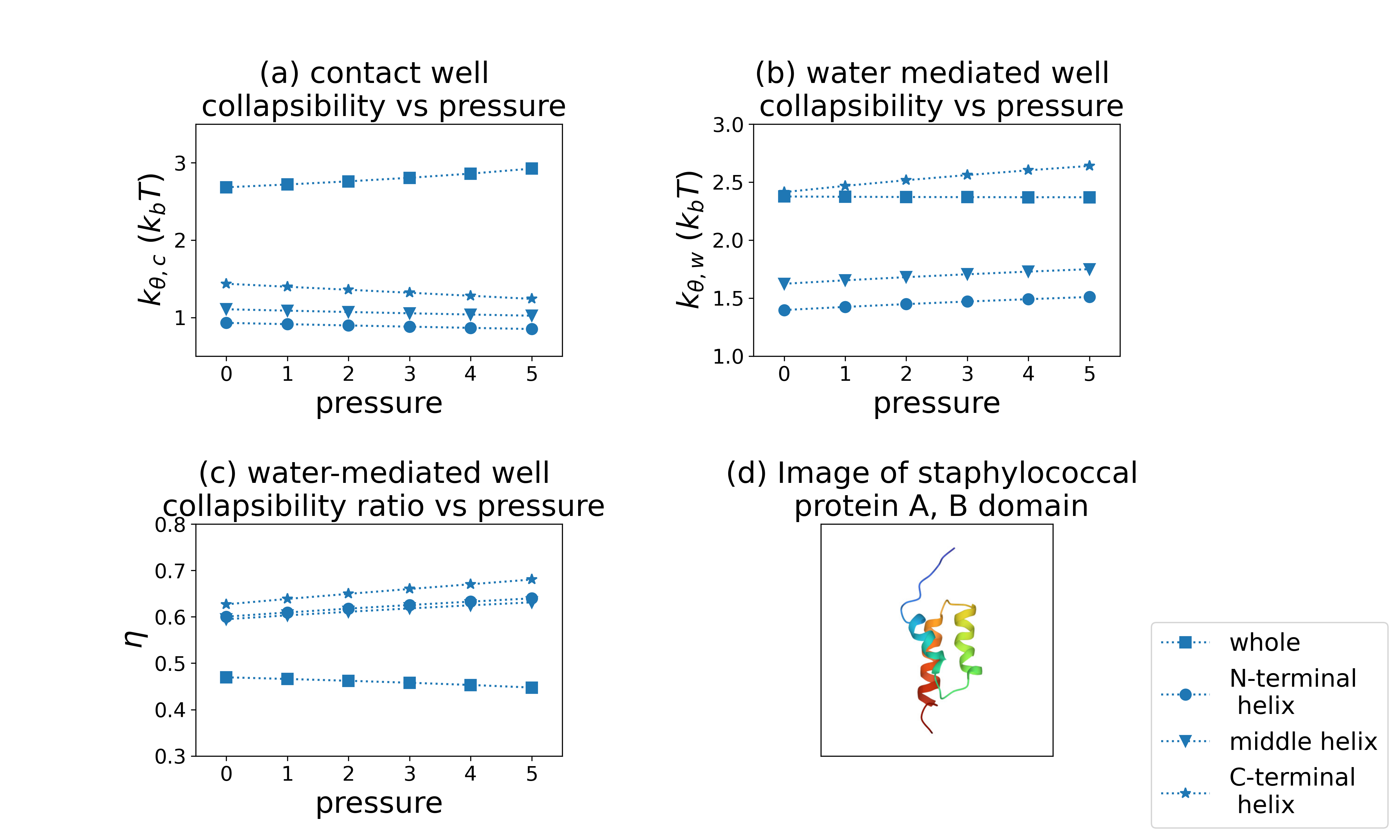}
\caption{\label{fig:1bdd} Pressure dependence of (a) contact-well collapsibility $k_{\theta,c}$, (b) water-mediated-well collapsibility $k_{\theta,w}$, and (c) water-mediated collapsibility ratio $\eta$ for the three $\alpha$-helical regions and the full sequence of the B domain of Staphylococcus aureus protein A (1BDD). (d) Three-dimensional structure of the protein, colored from the N-terminus (blue) to the C-terminus (red).}
\end{figure*}

\begin{figure*}
\centering
\captionsetup[subfigure]{labelformat=empty}
\includegraphics[width=\textwidth]{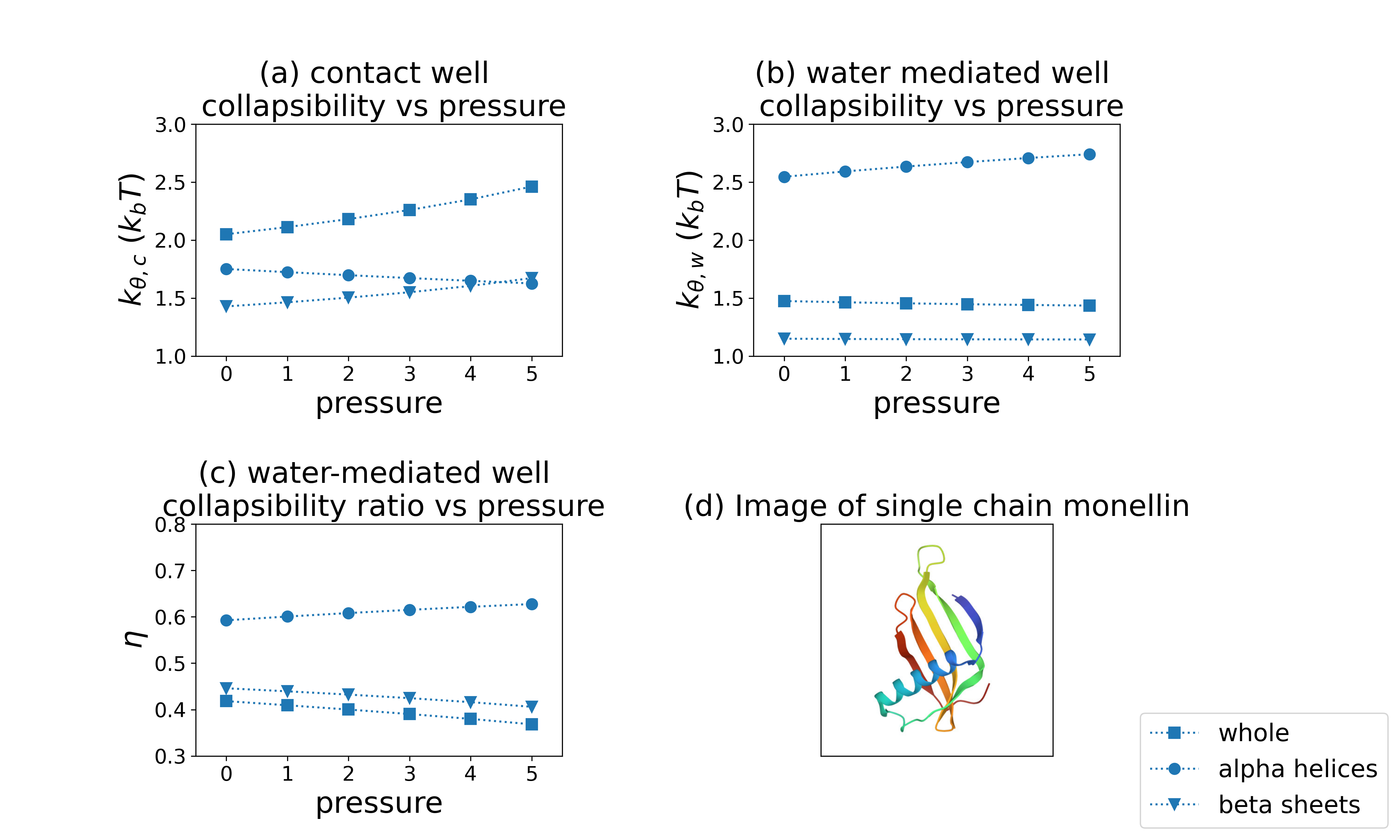}
\caption{\label{fig:1iv9} Pressure dependence of (a) contact-well collapsibility $k_{\theta,c}$, (b) water-mediated-well collapsibility $k_{\theta,w}$, and (c) water-mediated collapsibility ratio $\eta$ for the $\alpha$-helical and $\beta$-sheet regions and the full sequence of single-chain monellin (1IV9). (d) Three-dimensional structure of single-chain monellin.}
\end{figure*}

\subsubsection{Pressure-dependent collapsibility in the isolated helical region is distinct from the full-length sequence}
Figures~\ref{fig:1bdd} and \ref{fig:1iv9} reveal distinct pressure responses between the helical regions and their corresponding full-length proteins. From Figure ~\ref{fig:1bdd}, although the full sequences of protein A and single-chain monellin become less collapsible toward the contact well as pressure increases, their isolated $\alpha$-helical regions exhibit the opposite trend, becoming more collapsible toward direct-contact configurations.

The water-mediated response also differs between the helical regions and the full sequences. While the full-length proteins exhibit enhanced water-mediated collapsibility with increasing pressure, their $\alpha$-helical regions show a weaker tendency toward water-mediated collapse. This contrast suggests that interactions between structural regions, rather than interactions within individual helices alone, play an important role in determining the overall pressure response.

One possible explanation is that longer-range interactions between helices contribute to the formation of internal cavities, which may become hydrated under increasing pressure. These interactions are included in the full-length calculation but are absent when individual helical regions are analyzed separately. In single-chain monellin (Figure~\ref{fig:1iv9}), the $\beta$-sheet region exhibits pressure-dependent trends similar to those of the full sequence, further supporting the importance of non-local interactions in determining overall collapsibility.

\begin{figure*}
\centering
\captionsetup[subfigure]{labelformat=empty}
\includegraphics[width=\textwidth]{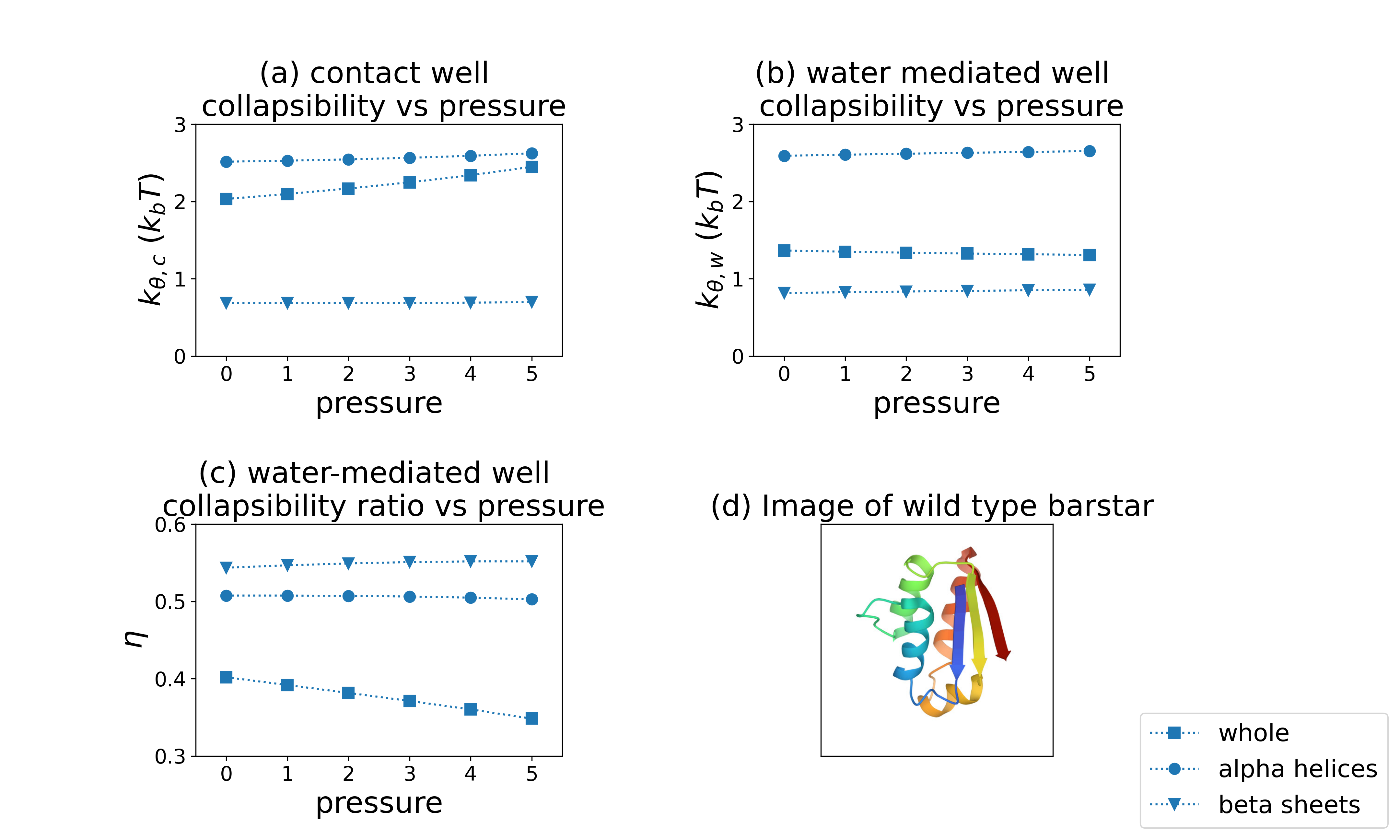}
\caption{\label{fig:1bta} Pressure dependence of (a) contact-well collapsibility $k_{\theta,c}$, (b) water-mediated-well collapsibility $k_{\theta,w}$, and (c) water-mediated collapsibility ratio $\eta$ for the $\alpha$-helical and $\beta$-sheet regions and the full sequence of wild-type barstar (1BTA). (d) Three-dimensional structure of wild-type barstar.}
\end{figure*}

\subsubsection{Arrangement of secondary structures on the sequence modulates pressure-dependent collapsibility}
Unlike protein A and single-chain monellin, wild-type barstar contains alternating $\alpha$-helical and $\beta$-sheet regions along its primary sequence. As shown in Figure~\ref{fig:1bta}, the pressure dependence of collapsibility in its individual secondary-structure regions does not exhibit a clear relationship with that of the full-length protein.

This observation suggests that the contribution of a particular secondary-structure type to overall collapsibility depends not only on its local geometry but also on how it is distributed along the sequence and connected through longer-range interactions. In proteins with interspersed secondary structures, the contributions of individual regions may be more strongly coupled, making their pressure responses less straightforward to separate.

\subsection{Pressure-dependence of collapsibility in multidomain proteins exhibits a similar trend as their individual domains}

To extend the regional analysis to larger proteins, we examined the pressure dependence of collapsibility in the two domains of 3-phosphoglycerate kinase (1QPG). The collapsibility metrics were calculated separately for each domain and compared with those of the full-length protein.

\begin{figure*}
\centering
\captionsetup[subfigure]{labelformat=empty}
\includegraphics[width=\textwidth]{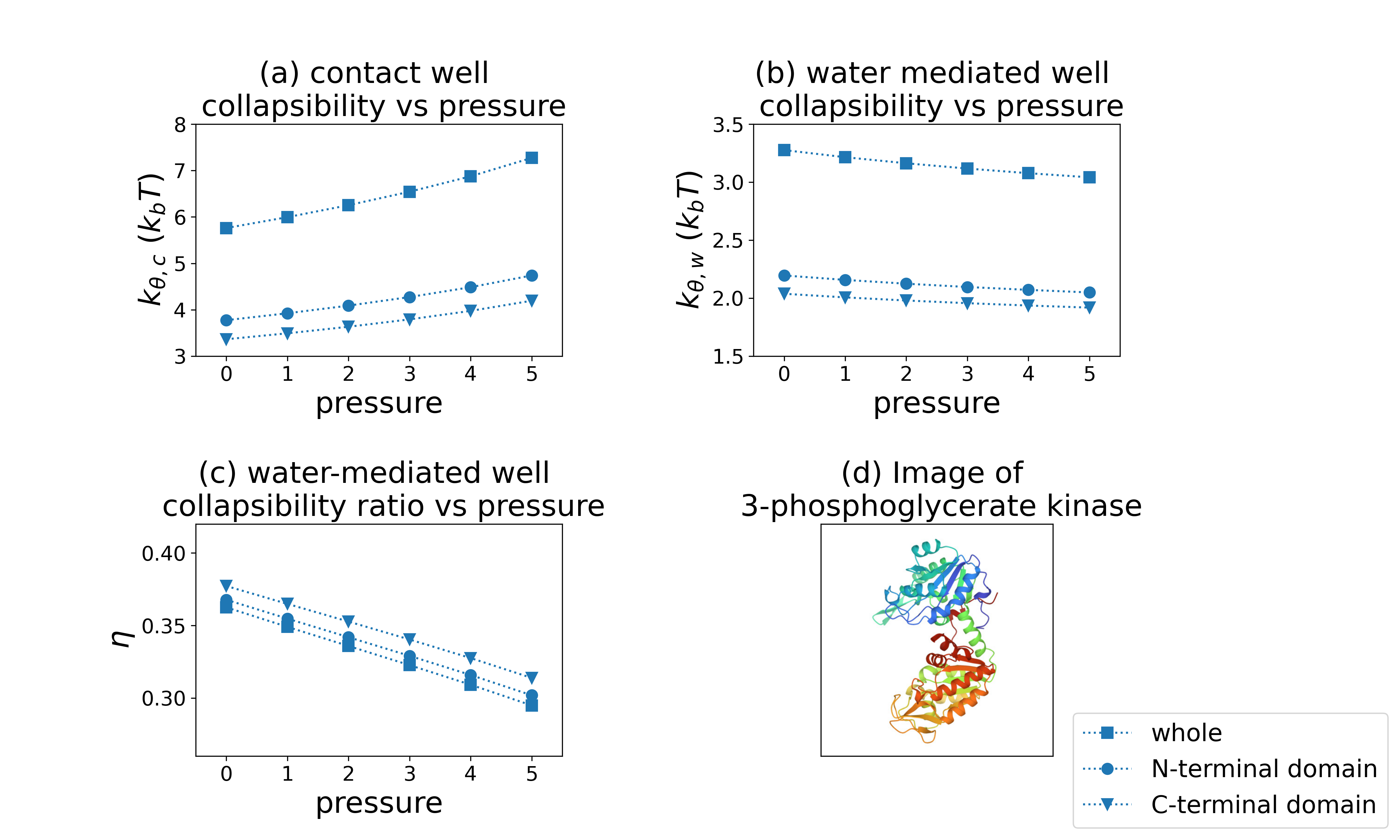}
\caption{\label{fig:pgk} Pressure dependence of (a) contact-well collapsibility $k_{\theta,c}$, (b) water-mediated-well collapsibility $k_{\theta,w}$, and (c) water-mediated collapsibility ratio $\eta$ for the two structural domains and the full sequence of 3-phosphoglycerate kinase (1QPG). (d) Three-dimensional structure of 3-phosphoglycerate kinase, colored from the N-terminus (blue) to the C-terminus (red).}
\end{figure*}

Figure~\ref{fig:pgk} shows that both the domains exhibit pressure-dependent collapsibility trends similar to those of the full-length protein. Unlike the individual $\alpha$-helical regions examined in protein A and single-chain monellin, neither domain shows a pronounced reversal in its pressure response relative to the complete sequence.

This similarity suggests that the two domains possess comparable intrinsic responses to pressure and that their individual contact networks capture important features of the full-length protein's collapsibility. The results are consistent with relatively independent domain-level contributions to pressure-dependent collapse, although interactions between domains may still influence the overall conformational response.

Together, the regional analyses demonstrate that pressure-dependent collapsibility is influenced by the hierarchical organization of protein structure. Local secondary-structure regions can exhibit pressure responses distinct from those of the full sequence, whereas larger structural domains may retain collapsibility trends similar to those of the complete protein. These findings highlight the importance of secondary and tertiary structures in determining how proteins respond to pressure.

\section{Discussion}
\subsection{Pressure denaturation depends on both secondary and tertiary structure}
In this study, we extended the analytical collapsibility framework proposed by  Samanta et al.~\cite{collapsibility} by incorporating pressure-dependent interactions that account for both direct-contact and water-mediated interactions. Our results show that, although increasing pressure destabilizes the native folded state, proteins generally retain or exhibit even enhanced collapsibility toward the water-mediated state. Furthermore, proteins with non-local native contacts tend to exhibit greater enhancement of the water-mediated collapsibility under pressure. In contrast, isolated $\alpha$-helical regions show the opposite behavior, becoming more collapsible toward the contact well as pressure increases.

These findings are consistent with the established mechanism of pressure-induced denaturation, in which increasing pressure promotes water penetration into internal cavities and disrupts native residue--residue contacts \cite{roy2024role,arsiccio2021pressure}. The greater pressure sensitivity of proteins with more non-local contacts suggests that long-range interactions may be particularly susceptible to pressure-induced hydration, whereas local interactions within $\alpha$-helices may be less affected. This interpretation is supported by all-atom simulations and recent experiments demonstrating pressure-induced stabilization of isolated $\alpha$-helices \cite{alpha_stability,makhatadze2025effects}, as well as theoretical studies showing that pressure favors the dissociation of coiled-coil structures while preserving the helicity of individual chains \cite{sumi2021water}. Consistent with these observations, isolated $\alpha$-helical regions and the single-chain retro-GCN4 leucine zipper (1C94) exhibit enhanced contact-well collapsibility and reduced water-mediated collapsibility with increasing pressure. Reduced water-mediated collapsibility is also observed in two other small, predominantly $\alpha$-helical proteins (1VII and 1ENH). In contrast, larger $\alpha$-helix-rich proteins, such as acetoacetyl-CoA thiolase (1QFL), exhibit enhanced water-mediated collapsibility, potentially due to long-range interhelical interactions and associated internal cavities. High-pressure NMR experiments further demonstrating that cavity position influences pressure-dependent conformational dynamics \cite{zhang2022high}. Together, these findings highlight the importance of tertiary packing and native contact architecture, rather than secondary-structure composition alone, in governing pressure-dependent collapsibility.

\subsection{Protein can obtain partially structured intermediates under pressure or chemical denaturation}
Our findings reveal that proteins retain their propensity for water-mediated collapse under elevated pressure, suggesting the persistence of compact intermediate states even when native-state folding is disfavored. This finding is consistent with numerous studies reporting metastable intermediates during protein collapse and folding \cite{bhatia2019observation, de2024mapping, agam2024folding}, as well as experimental observations of partially structured intermediates during pressure-induced unfolding\cite{CTL9, trypsin, myoglobin}. For example, Royer et al. \cite{CTL9} reported residue-specific pressure-unfolding profiles of CTL9 I98A that deviate from a simple two-state description. Ruan et al. \cite{trypsin} identified an intermediate state during pressure-induced unfolding of trypsin at approximately 6 kbar using fluorescence spectroscopy. Similarly, Heremans et al. \cite{myoglobin} showed, using FTIR measurements, that pressure unfolding preserves part of the $\alpha$-helical structure of myoglobin, whereas thermal unfolding causes extensive structural disruption. These observations suggest that pressure-induced denaturation preserves residual structure, consistent with our prediction that proteins may remain collapsible toward water-mediated configurations even when native contacts are destabilized.

Partially structured intermediates have also been observed during chemical denaturation \cite{hua_urea, jha_GdmCl_RnaseH, DMG_review,jha_GdmCl_MNEI}. One example is the dry molten globule (DMG), a partially expanded state in which the hydrophobic core remains relatively dehydrated. Hua et al. \cite{hua_urea} observed a DMG-like intermediate during lysozyme unfolding in 8 M urea, where urea penetrates the hydrophobic core before extensive water penetration. Similarly, DMG-like intermediates have been identified during GdmCl-induced unfolding of both single-chain monellin \cite{jha_GdmCl_MNEI} and \textit{Escherichia coli} ribonuclease HI \cite{jha_GdmCl_RnaseH}, with the latter study proposing a two-stage denaturation mechanism involving DMG formation followed by solvation of the protein interior.  Within our framework, proteins exhibiting reduced native-contact collapsibility without substantially enhanced water-mediated collapsibility correspond approximately to the first quadrant of Figure 4(a). These could serve as be candidates for DMG-like intermediates. This includes the small $\alpha$-helical proteins 1VII and 1ENH. However, most proteins fall in the fourth quadrant, exhibiting enhanced water-mediated collapsibility under pressure. This suggests that hydrated, partially collapsed conformations may be more broadly favored than DMG-like states during pressure denaturation, consistent with its water-penetration mechanism. Direct characterization of protein hydration would nevertheless be required to identify these intermediates.

\subsection{Application to other environmental conditions}
Our framework could also be extended to other denaturation conditions. Chemical denaturants destabilize proteins by altering solvent-mediated interactions or interacting directly with protein residues \cite{osterlund2025mass, nirmalraj2023situ}, which provides another approach to probe protein folding dynamics. Samiotakis and co-workers developed a urea-concentration-dependent energy function for coarse-grained simulations \cite{multiscale} and applied it to the folding dynamics of Trp-cage \cite{Trpcage_urea}. Incorporating similar concentration-dependent interactions into our framework could allow collapsibility to be studied under chemical denaturation and potentially provide insight into the formation of intermediates such as DMGs. In addition, cold denaturation could also be investigated by introducing temperature-dependent desolvation interactions, accounting for changes in protein hydration and solvent thermodynamics at low temperatures \cite{babu2004direct, arsiccio2020heightened}.

\section{Conclusion}
In conclusion, our results suggest that pressure-induced denaturation does not necessarily eliminate protein collapsibility but instead changes the relative tendency of proteins to collapse toward native-contact and water-mediated configurations. In other words, the propensity to collapse is encoded in natural sequences even at high pressures. However, the pressure response depends on the native topology. By incorporating effective pressure-dependent desolvation interactions, our theory provides a quantitative approach to understanding protein collapse under hydrostatic pressure and offers a basis for investigating other environmental perturbations.

\section{Limitations}
A limitation of the present framework is that pressure enters the collapsibility calculation in Eq.~(8) only through the pressure-dependent attractive Gaussian width $\sigma_\kappa(p)$, while the attractive amplitude is held constant. This parameterization is a modeling choice rather than a unique representation of pressure-dependent residue interactions. Furthermore, although the repulsive Gaussian amplitude $h_{\nu,w}(p)$ is fitted to reproduce the pressure-dependent desolvation barrier, it does not explicitly enter Eq.~(8). Consequently, the calculated collapsibility captures the effects of pressure-dependent changes in the effective attractive interaction range but does not directly account for changes in the desolvation barrier height. Future extensions could incorporate pressure dependence in both attractive and repulsive interactions to more fully characterize their contributions to protein collapsibility.

\appendix

\section{Theory of Collapsibility}
\label{app:collapsibility}

The interaction potential $\mathcal{V}(\mathbf{r}(s))$ introduced in Eq.~\ref{eq2} is given by

\begin{equation}
\begin{aligned}
\mathcal{V}(\mathbf{r}(s)) =
& \frac{\nu}{\left(2 \pi a_0^2\right)^{3/2}}
\sum_{s=0}^N \sum_{s^{\prime}=0}^N
\mathrm{e}^{-\frac{\left(\mathbf{r}(s)-\mathbf{r}(s^{\prime})\right)^2}{2 a_0^2}} \\
& -\frac{\kappa}{\left(2 \pi \sigma^2\right)^{3/2}}
\sum_{\left\{s_i, s_j\right\}}
\mathrm{e}^{-\frac{\left(\mathbf{r}(s_i)-\mathbf{r}(s_j)\right)^2}{2 \sigma^2}}
\end{aligned}
\label{eq:appendix_potential}
\end{equation}

\noindent
where $a_0$ is the diameter of an amino acid residue and $\sigma$ is the average distance between two residues that are in native contact in the database~\cite{collapsibility}. Each Gaussian is normalized and multiplied by a coefficient, with $\nu$ corresponds to the repulsive term and $\kappa$ corresponds to the attractive term, respectively, where $\nu$ is set to $\frac{4}{3}\pi a_0^3$. The widths of the Gaussians are chosen to approximate the effective interaction ranges. The repulsive term represents excluded-volume interactions between possible residue pairs, whereas the attractive term acts only between residue pairs that form native contacts.

In an earlier study, we used path-integral and variational approaches to predict the average size of a polymer chain described by the modified Edwards Hamiltonian~\cite{Rg_theory}. A virtual ideal chain was introduced with the same Hamiltonian but without the interaction potential. By comparing the Hamiltonians of the real polymer and the virtual ideal chain, one can calculate the correction to the chain's radius of gyration. For a real chain with monomer size $a_0$, an effective monomer size $a$ can be chosen for the ideal chain such that the first-order correction vanishes. Since the radius of gyration of an ideal chain with monomer size $a$ is given by $R_g^2=Na^2/6$~\cite{Rg_theory}, the radius of gyration of the real chain can then be estimated. Detailed calculations are provided elsewhere~\cite{collapsibility}. The resulting self-consistent equation relating $a$ to $a_0$ is

\begin{equation}
\begin{aligned}
\frac{1}{a_0^2}-\frac{1}{a^2}
= & \nu
\frac{
\left(\frac{3}{2}\right)^{5/2}
\left(\frac{\pi}{2}\right)^{3/2}
}{
\left(a^2\right)^{5/2}
N^{3/2}
\left(\sum_{n=1}^N \frac{1}{n^2}\right)
} \\
& \times
\sum_{s=0}^N \sum_{s^{\prime}=0}^N
\frac{
\sum_{n=1}^N
\frac{1-\cos\left[n\pi(s-s^{\prime})/N\right]}{n^4}
}{
\left(
\sum_{n=1}^N
\frac{1-\cos\left[n\pi(s-s^{\prime})/N\right]}{n^2}
+\frac{3\pi^2 a_0^2}{2a^2N}
\right)^{5/2}
} \\
& -\kappa
\frac{
\left(\frac{3}{2}\right)^{5/2}
\left(\frac{\pi}{2}\right)^{3/2}
}{
\left(a^2\right)^{5/2}
N^{3/2}
\left(\sum_{n=1}^N \frac{1}{n^2}\right)
} \\
& \times
\sum_{\left\{s_i,s_j\right\}}
\frac{
\sum_{n=1}^N
\frac{1-\cos\left[n\pi(s_i-s_j)/N\right]}{n^4}
}{
\left(
\sum_{n=1}^N
\frac{1-\cos\left[n\pi(s_i-s_j)/N\right]}{n^2}
+\frac{3\pi^2\sigma^2}{2a^2N}
\right)^{5/2}
}
\end{aligned}
\label{eq:appendix_monomer}
\end{equation}

The right-hand side of Eq.~\ref{eq:appendix_monomer} contains two contributions: the $\nu$-term and the $\kappa$-term, representing the effects of repulsive and attractive interactions on the radius of gyration, respectively.

Collapsibility $\kappa_\theta$ characterizes the attractive interaction strength required between native contact pairs to balance the excluded-volume effect. This balance occurs at the theta point, where attractive and repulsive contributions cancel. Thus, in Eq.~\ref{eq:appendix_monomer}, $\kappa_\theta$ can be obtained by setting $a=a_0$ and equating the $\nu$-term and the $\kappa$-term. The resulting expression is~\cite{collapsibility}

\begin{equation}
\begin{aligned}
\kappa_\theta
= \frac{4}{3}\pi a_0^3
\frac{
\displaystyle
\sum_{s=0}^N \sum_{s^{\prime}=0}^N
\frac{
\sum_{n=1}^N
\frac{1-\cos\left[n\pi(s-s^{\prime})/N\right]}{n^4}
}{
\left(
\sum_{n=1}^N
\frac{1-\cos\left[n\pi(s-s^{\prime})/N\right]}{n^2}
+\frac{3\pi^2}{2N}
\right)^{5/2}
}
}{
\displaystyle
\sum_{\left\{s_i,s_j\right\}}
\frac{
\sum_{n=1}^N
\frac{1-\cos\left[n\pi(s_i-s_j)/N\right]}{n^4}
}{
\left(
\sum_{n=1}^N
\frac{1-\cos\left[n\pi(s_i-s_j)/N\right]}{n^2}
+\frac{3\pi^2\sigma^2}{2a_0^2N}
\right)^{5/2}
}
}
\end{aligned}
\label{eq:appendix_collapsibility}
\end{equation}

A smaller value of $\kappa_\theta$ indicates that a protein sequence is more prone to collapse. Samanta et al.~\cite{collapsibility} showed that proteins with long-range native contacts, characteristic of $\beta$-sheet architectures, generally exhibit greater collapsibility.

\section{Pressure-Dependent Desolvation Potential}
\label{app:desolvation}

Pressure-dependent water-mediated desolvation potentials have been characterized in previous studies~\cite{hummer1998pressure,hillson1999pressure,Desolvation}. Unlike a single-well interaction potential, the desolvation potential contains a contact well and a water-mediated well separated by a desolvation barrier. The water-mediated well represents a local energy minimum in which a water molecule separates two residues, while the desolvation barrier represents the energy required to remove the intervening water molecule and form a direct native contact.

As pressure increases, the contact well becomes shallower while the desolvation barrier increases. In contrast, the depth of the water-mediated well remains approximately independent of pressure. Consequently, direct native contacts become less stable under increasing pressure, whereas water-mediated interactions retain their stability, with an increasing barrier separating the two interaction states.

In this study, we adopt the pressure-dependent parameters of the CHOGG desolvation model presented by Gasic and Cheung~\cite{Desolvation}:

\begin{equation}
\begin{aligned}
\epsilon(p) &= \epsilon_0-0.076p,\\
\epsilon'(p) &= 0.2,\\
\epsilon''(p) &= \epsilon_0''+0.127p.
\end{aligned}
\label{eq:appendix_desolvation}
\end{equation}

\noindent
Here, $\epsilon$ and $\epsilon'$ denote the absolute values of the contact-well and water-mediated-well depth, respectively, while $\epsilon''$ represents the desolvation barrier energy. The parameters $\epsilon_0$ and $\epsilon_0''$ specify the contact-well energy and desolvation barrier height at atmospheric pressure ($p=0$). All three energy quantities are expressed in kcal/mol. Pressure follows the convention of Gasic and Cheung~\cite{cheung2002protein, Desolvation}, where $p=0$ represents atmospheric pressure and one pressure unit corresponds to approximately 760 bar.

\begin{acknowledgments}
JW is grateful for the support of the National Science Foundation (MCB 2221824). AG was supported by the Mathematical and Physical Sciences Ascending Postdoctoral Research Fellowship (MPS-Ascend) from the National Science Foundation grant DMR 2137680. DT  was supported by the
NSF (grant CHE 2320256) and by the Welch
Foundation (grant F-0019 administered through the Collie-Welch Regents Chair).
This work was also
partially supported by the BRaVE project funded by the U.S. Department of Energy (DOE), Office of Science, Office of Biological and Environmental Research, under FWP 81832. Pacific Northwest National Laboratory is a multi-program national laboratory operated by Battelle for the DOE under Contract DE-AC05-76RL01830.
\end{acknowledgments}

\bibliographystyle{unsrt}
\bibliography{ref}
\end{document}